\documentclass[a4paper,12pt]{article}

\pdfoutput=1 

\usepackage[hmargin=.71in,vmargin=1.1in]{geometry}
\usepackage{indentfirst}
\usepackage{amsfonts}
\usepackage{mathrsfs}
\usepackage{amsmath}
\usepackage{mathabx}
\usepackage{amssymb}
\usepackage{authblk}
\usepackage{cite}
\usepackage{xcolor}
\usepackage{mathtools}
\usepackage{tensor}
\usepackage{physics}
\usepackage{graphicx}
\usepackage{bm}
\usepackage{upgreek}
\usepackage{braket}
\usepackage{color,soul}
\usepackage{csquotes}
\usepackage{caption}
\usepackage{subcaption}
\usepackage{slashed}
\usepackage{mathtools}

\usepackage[bookmarksnumbered=true,bookmarksopen=true]{hyperref}
 \hypersetup{colorlinks,
             linkcolor=[rgb]{0,0.3,0.6}, 
             citecolor=[rgb]{0,0.3,0.6}, 
             urlcolor=[rgb]{0,0.3,0.6}}

\newcommand\mi{\mathrm{i}}
\newcommand\me{\mathrm{e}}

\newcommand\pp{\uppi}
\newcommand{\dif}{\mathrm{d}}

\newcommand{\red}[1]{\textcolor{red}{#1}}

\DeclareMathOperator{\arctanh}{arctanh}

\usepackage{tikz}
\usetikzlibrary{matrix}

\begin{document}

\title{\Large\textbf{Axisymmetric generalization of 
zero-scalar-curvature  solutions from the Schwarzschild metric via the Newman-Janis algorithm}}

\author[a]{Chen Lan\thanks{stlanchen@126.com}}
\author[b,c]{Zi-Xiao Liu\thanks{liuzixiao@mail.nankai.edu.cn}}	
\author[b]{Yan-Gang Miao\thanks{Corresponding author: miaoyg@nankai.edu.cn.}}

\affil[a]{\normalsize{\em Department of Physics, Yantai University, 30 Qingquan Road, Yantai 264005, China}}
\affil[b]{\normalsize{\em School of Physics, Nankai University, 94 Weijin Road, Tianjin 300071, China}}
\affil[c]{\normalsize{\em National Supercomputer Center in Tianjin, Tianjin 300457, China}}

\date{ }

\maketitle

\begin{abstract}

We address a specific issue of the Newman-Janis algorithm: How to
determine the general form of the complex transformation for the Schwarzschild metric 
and ensure that the resulting axisymmetric metric satisfies the zero-scalar-curvature condition, $R=0$. 
In this context, the zero-scalar-curvature condition acts as a constraint.
Owing to this condition, we refer to the class of black holes 
as the ``Newman-Janis class of Schwarzschild black holes"
in order to emphasize Newman-Janis algorithm's potential as a classification tool for axisymmetric black holes.
The general complex transformation we derive not only generates 
the Kerr, Taub-NUT, and Kerr-Taub-NUT black holes under specific choices of parameters 
but also suggests the existence of additional axisymmetric black holes. 
Our findings open an alternative avenue using the Newman-Janis algorithm for the 
construction of new axisymmetric black holes.

\end{abstract}

\tableofcontents

\section{Introduction}
\label{sec:intr}

The Newman-Janis algorithm (NJA) \cite{Newman:1965tw} is a method 
used to derive axisymmetric black hole spacetimes from spherically symmetric ones. 
The term ``off-shell'' refers to the fact that these generated spacetimes do not necessarily 
satisfy the gravitational field equations \cite{Erbin:2016lzq}. This limitation has led to criticism because axisymmetric black holes obtained via the NJA often fail to satisfy these equations. 
For example, although the NJA can generate an axisymmetric black hole in Chern-Simons gravity, 
the Pontryagin density is non-zero, indicating a failure to satisfy the field equations \cite{Alexander:2009tp}.

The original NJA was formulated in terms of the Newman-Penrose formalism  \cite{Newman:1961qr}. 
It involves applying a complex transformation, followed by a change of coordinates or parameters, 
to derive a rotating solution \cite{Erbin:2014aya}. However, a modified version \cite{Azreg-Ainou:2014pra} 
of the NJA has been introduced, which eliminates the need for this additional complex coordinate transformation.
Its most notable success came from reproducing Kerr black holes\footnote{Taub-NUT black holes 
can also be retrieved from Schwarzschild black holes by the NJA \cite{Erbin:2016lzq}.} 
and discovering Kerr-Newman black holes 
in the Einstein-Maxwell theory \cite{Newman:1965my,Adamo:2014baa}. 
Given the observational relevance and diverse applications of rotating black holes \cite{Johannsen:2011dh,Bambi:2015kza,Konoplya:2016jvv,Arkani-Hamed:2019ymq,Perlick:2021aok}, the 
NJA has become a popular tool for constructing various types of axisymmetric black holes 
\cite{Modesto:2010rv,Bambi:2013ufa,Toshmatov:2017zpr,Kamenshchik:2023woo}. 
Despite this, the efforts continue \cite{Drake:1998gf,Beltracchi:2021ris,Lan:2023vaa} to understand the algorithm's underlying physical basis. 
As time goes on, the NJA has evolved into two distinct branches: 
one that retains the Newman-Penrose formalism \cite{Gurses:1975vu,Azreg-Ainou:2014pra,Dymnikova:2015hka} 
and the other that no longer relies on tetrads \cite{Erbin:2014aya,Erbin:2014lwa}.

In this work, we apply the NJA within the Newman-Penrose formalism 
but omit the complexification of variables and parameters. 
Our goal is to derive a general complex transformation for Schwarzschild black holes 
that yields axisymmetric black holes with zero-scalar-curvature metrics.\footnote{The zero-scalar-curvature condition  indicates either a vacuum solution or a spacetime generated by matter with a zero trace of energy-momentum tensors.
}
In other words, we aim to explore how many zero-scalar-curvature and axisymmetric black holes  
can be generated from Schwarzschild black holes by using the NJA. 
We refer to these solutions as the Newman-Janis (NJ) class of Schwarzschild black holes, 
emphasizing NJA’s potential as a classification tool for axisymmetric black holes. 
In general, the black holes derived from the same seed solution should belong to one NJ class because
they share the key physical properties \cite{Lan:2023vaa}.

The remainder of the paper is structured as follows.
In Sec.\ \ref{sec:taub-nut}, we begin by applying a modified version 
of the NJA \cite{Azreg-Ainou:2014pra} 
to retrieve Taub-NUT black holes from the Schwarzschild seed metric. 
We emphasize the function freedoms in the process of retrieving Taub-NUT black holes, 
which motivates a generalization of the modified NJA in Sec. \ref{sec:general}, 
where we also present a formal metric for axisymmetric black holes 
resulting from general complex transformations.
In Sec. \ref{sec:NJ-class}, we focus on the zero-scalar-curvature condition 
and derive the explicit form of the complex transformations introduced 
in the previous section. Sec.~\ref{sec:new} explores a new type 
of axisymmetric black holes predicted using our treatment. 
The final section, Sec.~\ref{sec:conclusion}, presents our conclusion, followed by two appendices detailing 
the specific forms of two curvature invariants.

\section{Non-complexified NJA for Taub-NUT black holes}
\label{sec:taub-nut}

In this section, we introduce the approach used~\cite{Azreg-Ainou:2014pra} for Kerr black holes 
and extend it to Taub-NUT black holes, aiming to further generalize 
this approach in the subsequent section. The term ``non-complexified" refers to 
the fact that this approach does not require extending coordinates and parameters into the complex domain.
We begin with the seed metric for a spherically symmetric spacetime,
\begin{equation}
\label{eq:seed}
\dif s^2 = -g(r) \dif t^2 + \frac{\dif r^2}{f(r)} + h^2(r) \dif\Omega^2_2,
\end{equation}
where $\dif\Omega^2_2 = \dif \theta^2 + \sin^2\theta \dif\phi^2$ is the metric of a unit two-sphere.

The first step is to rewrite this metric in the Newman-Penrose tetrad formalism. 
We introduce the advanced null coordinate,
\begin{equation}
\dif u = \dif t - \frac{\dif r}{\sqrt{g(r) f(r)}},
\end{equation}
which transforms the above metric to
\begin{equation}
\label{eq:adv-null-met}
\dif s^2 = -g(r) \dif u^2 - 2\sqrt{\frac{g(r)}{f(r)}} \dif u \dif r + h^2(r) \dif\Omega^2_2.
\end{equation}
This can then be expressed in terms of the Newman-Penrose tetrad as
\begin{equation}
\label{eq:metric-np}
g^{\mu\nu} = l^\mu n^\nu + l^\nu n^\mu - m^\mu \bar{m}^\nu - m^\nu \bar{m}^\mu,
\end{equation}
where $ l^\mu$, $n^\mu$, and $m^\mu$ are null vectors, and $\bar{m}^\mu$ 
is the complex conjugate of $m^\mu$. The explicit form of the tetrad vectors is given by
\begin{equation}
\label{eq:org-tetrad}
l^\mu = \delta^\mu_r, \qquad 
n^\mu =\sqrt{\frac{f}{g}}\delta^\mu_u - \frac{f}{2}\delta^\mu_r, \qquad 
m^\mu = \frac{1}{\sqrt{2} h} \left( \delta^\mu_\theta + \frac{\mi}{\sin\theta} \delta^\mu_\phi \right).
\end{equation}

The next step is our proposal: We generalize the functions $f(r)$, $g(r)$, and $h(r)$ to $F(r, N)$, $G(r, N)$, 
and $H(r, N)$, ensuring that they reduce to their original forms when the NUT charge $N$ vanishes:
\begin{equation}
\label{eq:condition}
\lim_{N\to 0} \{ F(r, N), G(r, N), H(r, N) \} = \{ f(r), g(r), h(r) \}.
\end{equation}
This relation serves as a boundary condition for the subsequent calculations.
We then introduce a complex coordinate transformation,\footnote{The term ``non-complexified'', as used in this paper, follows the convention introduced in Ref.\ \cite{Azreg-Ainou:2014pra}.
In that context, the complexification refers to the process of extending both the radial coordinate $r$ and physical parameters into the complex domain. 
In contrast, the non-complexification means that neither $r$ nor the parameters, such as mass $M$ or NUT charge $N$, are treated as complex. 
Nevertheless, a complex coordinate transformation is still applied. 
For instance, in deriving the Taub–NUT black hole from the Schwarzschild solution, one performs the transformation given in Eq.\ \eqref{eq:trans-nut} without complexifying $r$, $M$, or $N$. That is, this coordinate transformations is made in the complex domain, but the coordinate and the parameters themselves are real.}
\begin{equation}
\label{eq:trans-nut}
u \to u - 2\mi N \ln \sin \theta, \qquad r \to r - \mi N,
\end{equation}
substitute it into Eq.\ \eqref{eq:org-tetrad}, and obtain the following tetrad:
\begin{equation}
\label{eq:new-tetrad}
l^\mu = \delta^\mu_r, \qquad n^\mu = \sqrt{\frac{F}{G}}\delta^\mu_u - \frac{F}{2} \delta^\mu_r, \qquad 
m^\mu = \frac{1}{\sqrt{2} H} \left(- 2 \mi N \cot\theta \delta^\mu_u + \delta^\mu_\theta + \frac{\mi}{\sin\theta}\delta^\mu_\phi \right).
\end{equation}
Using this tetrad and Eq.\ \eqref{eq:metric-np}, we generalize the metric as follows:
\begin{equation}
\dif s^2 = -2\sqrt{\frac{G}{F}} \left[\dif u \dif r +2 N \cos \theta \dif\phi \dif r \right] - G[\dif u + 2 N \cos \theta \dif\phi]^2 + H^2 \left(\dif\theta^2 + \sin^2\theta \dif\phi^2 \right).\label{gTNN}
\end{equation}
The above treatment differs from the traditional one applied to Taub-NUT black holes in Ref.~\cite{Erbin:2016lzq} 
where the mass parameter is also transformed by $M \to M' = M - \mi N$.

To express the above metric in the Boyer–Lindquist coordinates, we perform the following coordinate transformation:
\begin{equation}
\label{eq:bl-transform}
\dif u \to \dif t + \lambda (r, N) \dif r, \qquad 
\dif \phi \to \dif \phi + \chi (r, N) \dif r.
\end{equation}
After considering the conditions $g_{tr} = g_{r\phi} = 0$, we derive
\begin{equation}
\label{eq:det-lambda}
\lambda = -\frac{1}{\sqrt{F G}}, \qquad \chi = 0,
\end{equation}
which are required by the axisymmetric and stationary preconditions. 
Subsequently, using the boundary condition given by Eq.\ \eqref{eq:condition}, we obtain
\begin{equation}
\lim_{N\to 0}  \lambda = - \lim_{N\to 0} \frac{1}{\sqrt{F G}} = - \frac{1}{\sqrt{f g}}.\label{limpara}
\end{equation}

For Schwarzschild black holes, where $f = g = 1 - 2M/r $, we find
\begin{equation}
\lim_{N \to 0} \frac{1}{\sqrt{FG}} = \frac{1}{f} = \frac{r}{r - 2M}.
\end{equation}
Following the procedure outlined in Ref.\ \cite{Azreg-Ainou:2014pra} for the Kerr case, we choose for the Taub-NUT case,
\begin{equation}
\label{eq:gauge-taub-nut}
\lambda = \frac{r^2 + N^2}{N^2 + 2M r - r^2},
\end{equation}
and solve $ F $ from Eq.\ \eqref{eq:det-lambda}:
\begin{equation}
F = \frac{(N^2 + 2M r - r^2)^2}{(r^2 + N^2)^2 G}.
\end{equation}
We note that the choice of $\lambda$, Eq.~\eqref{eq:gauge-taub-nut}, is based on our experience and familiarity with the Taub-NUT black hole metric. 
Later, we will show how the 
equations of motion, $ R^\mu_{\;\;\nu} = 0 $, 
along with the boundary conditions Eq.~\eqref{eq:condition}, can be used to determine the exact forms of the functions $F(r, N)$, $G(r, N)$, 
and $H(r, N)$.

By setting $ F = G $, we obtain
\begin{equation}
G = F = \frac{r^2 - 2M r - N^2}{r^2 +N^2}.
\end{equation}
The choice of $ F = G $ results from the characteristics of the seed metric, i.e., $ f = g $ in the Schwarzschild case.

Finally, by fixing $ H = \sqrt{r^2 + N^2} $, we recover the Taub-NUT black hole metric \cite{Griffiths:2009dfa},
\begin{equation}
\label{eq:metric-Taub-NUT}
\dif s^2 = -F\left(\dif t + 2N \cos{\theta}\dif\phi\right)^2 + F^{-1}\dif r^2 + (r^2 + N^2)\sin^2{\theta} \dif\phi^2 + \Sigma \dif\theta^2,
\end{equation}
where $\Sigma = r^2 + N^2$.
The choice of $ H $, similar to that of $ \lambda$, will be discussed in further detail. 
Before proceeding, we offer some remarks on the overall process.

In Eq.\ \eqref{eq:det-lambda}, the equation $\chi = 0$ reduces the number of free functions 
compared with the case of Kerr black holes \cite{Azreg-Ainou:2014pra}, but this leaves $H$ undetermined, 
requiring it to be manually fixed. In Table~\ref{tab:freedom}, we compare the function freedoms between Taub-NUT and Kerr black holes in the framework of the NJA.

\begin{table}[htp!]
\begin{center}
\begin{tabular}{c | c c c c c c c c}
\hline\hline
 & \em f & \em g & \em h & \em F & \em G & \red{\em H} & $\lambda$ & $\red{\chi}$ \\
 \hline
Taub-NUT BHs & giv & giv & giv  & det & det & \red{semi} & semi & \red{det}  \\
Kerr BHs & giv & giv & giv  & det & det & \red{det} & semi & \red{semi} \\
\hline\hline 
\end{tabular}
\end{center}
\caption{Function freedoms. }
\label{tab:freedom}
\end{table}

In Table \ref{tab:freedom},  ``giv" denotes ``given," meaning that the function is initially provided as a condition. 
``det" refers to ``determined," indicating that the function can be formally calculated 
if other functions are known.  ``semi" denotes ``semi-determined," 
meaning that the asymptotic behavior of the function with respect to 
a certain parameter is known, but its explicit form is inferred from the target metric. 
The table shows that, for the NJ class of Schwarzschild black holes, 
only two free functions must be fixed based on experience or unknown conditions. 
When $\chi$ is fixed to zero, the remaining function $H$ becomes free for Taub-NUT black holes. In contrast, $\chi$ becomes free for Kerr black holes when $H$ is fixed.

The presence of two semi-determined functions in the NJA suggests 
introducing additional conditions to fully determine them. 
The equations of motion provide a natural choice for this purpose. 
According to it, the NJA can partially circumvent the traditional challenge of being considered to be an {\em off-shell} method.
We begin with the metric in the Boyer–Lindquist coordinates,
\begin{equation}
\dif s^2 = -F \left[\dif t + 2N \cos \theta \dif \phi \right]^2 
+ \frac{\dif r^2}{F} + H^2 \left[\dif \theta^2 
+ \sin^2 \theta \dif \phi^2 \right],
\end{equation}
where we have used $ F = G $ and replaced $\lambda$ and $\chi$ by
\begin{equation}
\lambda \to -\frac{1}{F}, \qquad \chi \to 0
\end{equation}
in Eq.~\eqref{gTNN} under the coordinate transformation from Eq.~\eqref{eq:bl-transform}. 
Our goal is to use the vacuum equations of motion, $R^\mu_{\;\nu} = 0 $, along with the boundary conditions from Eq.\ \eqref{eq:condition}, to provide the explicit forms of $F$ and $H$.
In Sec.\ \ref{sec:NJ-class}, when we address the general case, 
we extend the vacuum equations discussed here to encompass the zero-scalar-curvature condition.

We begin by noting $ R^2_{\;2} = R^3_{\;3} $, 
which reduces the equations of motion to the following three independent equations:
\begin{subequations}
\label{eq:eom}
\begin{equation}
\label{eq:key1}
H^4 F'' + 2 H^3 F' H' + 4 N^2 F = 0,
\end{equation}
\begin{equation}
\label{eq:step1}
H F'' + 2 F' H' + 4 F H'' = 0,
\end{equation}
\begin{equation}
\label{eq:constants}
H^2 \left[H \left(F' H' + F H''\right) + F (H')^2 \right] = 2 N^2 F + H^2.
\end{equation}
\end{subequations}
From Eq.\ \eqref{eq:key1}, we derive the solution for $ F $,
\begin{equation}
\label{eq:f-sol}
F = \frac{\sqrt{c_1}}{2N} \sin\left[2N \left( \int \frac{\dif r}{H^2} + c_2 \right)\right],
\end{equation}
where $ c_1 $ and $ c_2 $ are constants of integration.

Substituting this expression for $ F $ into Eq.\ \eqref{eq:step1}, we arrive at the equation
\begin{equation}
H^3 H'' - N^2 = 0,
\end{equation}
which leads to the solution for $ H $,
\begin{equation}
\label{eq:solh}
H^2 = \frac{N^2}{c_3} + c_3 (r + c_4)^2,
\end{equation}
where $ c_3 $ and $ c_4 $ are also integration constants.
By substituting this result for $ H $ back into the expression Eq.\ \eqref{eq:f-sol} for $ F $, we obtain
\begin{equation}
\label{eq:solf}
F = \frac{\sqrt{c_1}}{2N} \sin\left[2 \left(\arctan\left(\frac{c_3 (r + c_4)}{N}\right) + c_2 N \right)\right].
\end{equation}
Finally, substituting these results into the third equation, Eq.\ \eqref{eq:constants}, 
enables us to determine the integration constants through the following relation:
\begin{equation}
\label{eq:solconsts}
2 N + c_3 \sqrt{c_1} \sin(2 c_2 N) = 0.
\end{equation}
Thus, Eqs.\ \eqref{eq:solh}, \eqref{eq:solf}, and \eqref{eq:solconsts} provide the general solutions to the equations of motion, Eq.\ \eqref{eq:eom}.

We now proceed to determine the constants of integration using the conditions provided in Eq.\ \eqref{eq:condition}. 
First, applying the condition
\begin{equation}
\lim_{N \to 0} H^2 = c_3 (r + c_4)^2 = r^2,
\end{equation}
we find
\begin{equation}
c_3 = 1, \qquad c_4 = 0.
\end{equation}
Next, we derive $ c_2 $ from Eq.\ \eqref{eq:solconsts},
\begin{equation}
c_2 = \frac{\pi + \arcsin\left(\frac{2N}{\sqrt{c_1}}\right)}{2N},
\end{equation}
where we have omitted the periodicity inherent in the trigonometric function.
Substituting $ c_2 $, $ c_3 $, and $ c_4 $ into Eq.~\eqref{eq:solf}, we obtain
\begin{equation}
\label{eq:almostf}
F = \frac{-N^2 + r \left(r - \sqrt{-4N^2 + c_1}\right)}{N^2 + r^2}.
\end{equation}
Next, applying the boundary condition
\begin{equation}
\lim_{N \to 0} F = 1 - \frac{\sqrt{c_1}}{r} = 1 - \frac{2M}{r},
\end{equation}
we deduce that the parameter $ c_1 $ must be $4M^2$ plus a $N$-dependent function that vanishes as $N\to 0$. 
Considering the balance of dimensions, we arrive at 
\begin{equation}
c_1 = 4(M^2 + N^2).
\end{equation}
Therefore, we have retrieved the metric of Taub-NUT black holes without arbitrarily assigning specific forms to any function at any stage of the above derivation.

Our new treatment demonstrates that, starting from the Schwarzschild metric and applying the complex transformation in Eq.\ \eqref{eq:trans-nut}, the Taub-NUT metric is the unique solution that satisfies the vacuum equations of motion $R_{\mu\nu}=0$.

\section{General NJA without complexifications}
\label{sec:general}

In this section, we derive the forms of $\lambda$ and $\chi$ using a kind of matching conditions\footnote{It is Eq.~\eqref{eq:matching}, a generalized boundary condition similar to Eq.~\eqref{eq:condition}.} instead of specifying the initial transformations. 
Our goal is to generalize the NJA for Schwarzschild black holes by introducing a broader class of transformations.

To extend our treatment from the previous section to arbitrary transformations, we consider a generalized complex transformation of Eq.\ \eqref{eq:trans-nut}:
\begin{equation}
u \to u - \mi \alpha(\theta, p), \qquad r \to r + \mi \beta(\theta, p),\label{gcomtrans}
\end{equation}
where the functions $\alpha(\theta, p)$ and $\beta(\theta, p)$ depend on certain parameters that are
denoted collectively by $p$, such as the NUT charge in Eq.~\eqref{eq:trans-nut}. 
For simplicity, we use $p$ to represent any relevant parameter and use the symbols $\alpha(\theta)$ and $\beta(\theta)$ but keep in mind that $\alpha$ and $\beta$ depend on $p$.

First, we extend the functions in the seed metric, 
originally appearing in the advanced null coordinate in Eq.\ \eqref{eq:adv-null-met},
and derive the following null tetrad by substituting Eq.\ \eqref{gcomtrans} into Eq.\ \eqref{eq:org-tetrad}:
\begin{equation}
l^\mu = \delta^\mu_r, \qquad
n^\mu = \sqrt{\frac{F}{G}} \delta^\mu_u - \frac{F}{2} \delta^\mu_r,
\qquad
m^\mu = \frac{1}{\sqrt{2} H} \left[- \mi \alpha'(\theta) \delta^\mu_u + \mi \beta'(\theta) \delta^\mu_r + \delta^\mu_\theta + \frac{\mi}{\sin\theta} \delta^\mu_\phi \right],\label{gertetrad}
\end{equation}
where the prime denotes the derivative with respect to $\theta$ and, the functions $F(r, \theta, p)$, $G(r, \theta, p)$, and $H(r, \theta, p)$ depend on the parameter $p$ and satisfy the matching condition
\begin{equation}
\label{eq:matching}
\lim_{p \to 0} \{F(r, \theta, p), G(r, \theta, p), H(r, \theta, p)\} = \{f(r), g(r), h(r)\}.
\end{equation}
This condition must be satisfied because, 
for a given transformation Eq.\ \eqref{gcomtrans}, 
the axisymmetric black holes produced by the NJA
related to the functions $ G $, $ F $, and $ H $
reduce to the original spherically symmetric metric Eq.\ \eqref{eq:seed} 
related to $ f $, $ g $, and $ h $, as $p$ approaches zero.
Following the procedure for deriving Eq.~\eqref{gTNN} but replacing  Eqs.~\eqref{eq:trans-nut} and \eqref{eq:new-tetrad} by
Eqs.~\eqref{gcomtrans} and \eqref{gertetrad}, respectively, we provide the following metric:
\begin{equation}
\label{denmetricBL}
\begin{split}
\dif s^2 =  -G& \left[\dif u + \alpha'(\theta) \sin \theta \dif \phi \right]^2 
+ H^2 \left(\dif \theta^2 + \sin^2 \theta \dif \phi^2 \right) \\
& + 2 \sqrt{\frac{G}{F}} \left[\beta'(\theta) \sin \theta \dif \phi - \dif r \right] \left[\dif u 
+\alpha'(\theta) \sin \theta  \dif \phi \right].
\end{split}
\end{equation}

Secondly, when the transformation Eq.\ \eqref{eq:bl-transform} is applied with the replacement of $N$ by $p$, 
the conditions $g_{tr} = g_{r\phi} = 0$ lead to the following solutions\footnote{Because $\lambda(r, p)$ and $\chi(r, p)$ originate from the Boyer–Lindquist coordinates, they do not depend on $\theta$. That is, the $\theta$ variable must be canceled out in the solutions.} for $\lambda(r, p)$ and $\chi(r, p)$:
\begin{equation}
\label{eq:solution}
\lambda (r, p) =  -\frac{H^2 \sqrt{{F}/{G}} + \alpha'(\theta) \beta'(\theta)}{F H^2 + (\beta'(\theta))^2},\qquad
\chi (r, p) =  \frac{\csc(\theta) \beta'(\theta)}{F H^2 + (\beta'(\theta))^2}.
\end{equation}
Following the reasoning stated in the previous section (see the contexts from  Eqs.~\eqref{eq:bl-transform}-\eqref{limpara} and the relevant analysis in Ref.\ \cite{Azreg-Ainou:2014pra}), 
we propose the following forms of $\lambda(r)$ and $\chi(r)$ as ${p \to 0}$:
\begin{equation}
\label{eq:guage}
\lambda (r) =  -\frac{h^2 \sqrt{{f}/{g}} + d}{fh^2 + b^2},\qquad
\chi (r) =- \frac{c}{fh^2 + b^2},
\end{equation}
where $b$, $c$, and $d$ are free parameters. Before continuing calculations, the following two points deserve attention:
\begin{itemize}
\item Only one of $\alpha'(\theta)$ and $\beta'(\theta)$ can be zero. If both are zero, no modification occurs to the seed metric or the null tetrad. If $\alpha'(\theta)$ is zero and $F = G$, both $F$ and $H$ can be determined, but if $\beta'(\theta)$ is zero, only $F$ can be determined, leaving $H$ undetermined. This indicates that it may not fix all free functions to use the equations of motion alone. Therefore, we will first examine the case of $\beta'(\theta) \neq 0$.
\item The assumption in Eq.\ \eqref{eq:guage} is motivated by the form of Eq.\ \eqref{eq:solution} and the matching condition in Eq.\ \eqref{eq:matching}, which is quite general. However, for a specific case, such as Taub-NUT black holes, the formulation Eq.~\eqref{eq:gauge-taub-nut} is inferred directly from the Taub-NUT metric in Eq.~\eqref{eq:metric-Taub-NUT}.
\end{itemize}

Finally, when $f = g $ in the seed metric, we can reasonably assume $F(r, \theta) = G(r, \theta) $. 
From Eqs.~\eqref{eq:solution} and \eqref{eq:guage}, we can express $F(r, \theta) $ as
\begin{equation}
F(r, \theta) = \frac{b^2 + c \beta'(\theta) \sin\theta + f(r) h^2(r)}{d + c \alpha'(\theta) \sin\theta + h^2(r) \sqrt{{f(r)}/{g(r)}}},
\end{equation}
and determine the function $H(r, \theta) $ owing to $\beta'(\theta) \neq 0$:
\begin{equation}
H^2(r, \theta) = -\frac{\beta'(\theta)}{c \sin\theta} \left[d + c \alpha'(\theta) \sin\theta + h^2(r) \sqrt{\frac{f(r)}{g(r)}}\right].
\end{equation}

For the Schwarzschild metric where $f = g = 1 - 2M/r $ and $h = r $, the resulting metric components of  Eq.~\eqref{denmetricBL} take the forms
\begin{subequations}
 \label{eq:metric-general}
\begin{equation}
g_{00} = -\frac{b^2 + c \beta' \sin \theta - 2Mr + r^2}{d + c \alpha' \sin \theta + r^2}, \qquad
g_{11} = \frac{d + c \alpha' \sin \theta + r^2}{b^2 - 2Mr + r^2},
\end{equation}
\begin{equation}
g_{03} = \frac{\sin \theta \left[\left(d + r^2\right) \beta' - \left(b^2 + r(r - 2M)\right) \alpha'\right]}{d + c \alpha' \sin \theta + r^2},
\end{equation}
\begin{equation}
\label{eq:problem}
g_{22} = -\frac{1}{c} \beta' \left[\left(d + r^2\right) \csc \theta + c \alpha'\right],
\end{equation}
\begin{equation}
g_{33} = \frac{\left[-c\left(b^2 + r(r - 2M)\right) (\alpha')^2 \sin \theta 
- \left(d + r^2\right)^2 \beta'\right] \sin \theta}{c\left(d + c \alpha' \sin \theta + r^2\right)}.
\end{equation}
\end{subequations}
If the metric described by the above components satisfies the condition of a vanishing Ricci curvature, 
it belongs to the NJ class of Schwarzschild black holes, which is discussed in the next section.

The spacetime produced by the NJA is asymptotically flat as long as $\beta'$ does not vanish. 
As $r \to \infty $, the Ricci curvature scalar behaves as
\begin{equation}
R \sim -\frac{2}{r^2} + \frac{c \sin \theta }{r^2 \beta '} + \frac{c (\beta'')^2 \sin \theta}{r^2 (\beta')^3} - \frac{c \left(\beta^{'''} \sin \theta + \beta'' \cos \theta\right)}{r^2 (\beta')^2} + O\left(r^{-3}\right).
\end{equation}
Furthermore, the formulation of Ricci scalar $R$ and Kretschmann scalar $K\equiv R^{\mu}_{\nu\alpha\beta}R_{\mu}^{\nu\alpha\beta}$ contains the denominator $8 [\beta'(\theta )]^3 \left[c \alpha '(\theta )\sin\theta +d+r^2\right]^3$; thus a potential curvature singularity exists at $r_s $ given by
\begin{equation}
r_s = \sqrt{- c \alpha' \sin\theta - d}.
\end{equation}
If $r_s $ is not real for all values of $\theta$, the metric describes a regular black hole spacetime. 
The event horizon is determined by the singularity of $g_{11} $,
\begin{equation}
b^2 - 2Mr + r^2 = 0,
\end{equation}
which is analogous to the horizon of Kerr black holes.

\section{NJ class of Schwarzschild black holes}
\label{sec:NJ-class}

In this section, we aim to explore whether it can yield additional zero-scalar-curvature black hole solutions by applying the NJA to Schwarzschild black holes. 
Specifically, we seek the general forms of the functions $\alpha(\theta)$ and $\beta(\theta)$ that satisfy the condition $R = 0$, indicating the zero scalar-curvature.\footnote{The vacuum solution from $R_{\mu\nu}=0$ satisfies the zero-scalar-curvature condition, and the spacetime generated by matter with a vanishing trace of the energy-momentum tensor, $T^\mu_{\;\;\mu}=0$, also satisfies this condition.
In other words, the zero-scalar-curvature condition encompasses a broader range of solutions than the vacuum equations of motion $R_{\mu\nu}=0$.} 
We note that the zero scalar-curvature is a less stringent condition than the vacuum Einstein equations, $R^\mu_{\;\nu} = 0$. 
Although any metric that satisfies the vacuum Einstein equations must be zero-scalar-curvature, 
a zero-scalar-curvature metric may not satisfy the Einstein equations. 
A zero-scalar-curvature metric can correspond to the scenarios involving a traceless energy-momentum tensor, $\Tr T^{\mu}_{\;\;\nu}=0$.
An additional issue arises concerning the interpretation of the NJA. 
Traditionally, the NJA has been understood primarily as a method for introducing rotation into a static black hole solution. 
However, it can also alter other physical properties, such as generating a nonzero NUT charge. 
The observation that the energy-momentum tensor has zero trace may evoke similarities with electrodynamics and prompt the question of whether the NJA could similarly introduce electric charge into an initially uncharged BH. 
From the perspective of symmetry, if the seed solution possesses an axisymmetric distribution of electric or magnetic charge, our construction method does not exclude the possibility of such charges emerging in our new solution.
Therefore, our idea is applicable to searching for new axisymmetric black holes from a spherically symmetric seed black hole in gravity coupled to other fields.

To proceed, we express the Ricci scalar $R$ as a polynomial in $r$:
\begin{equation}
\label{eq:ricci}
R = \frac{1}{R_d}\left(R_{n,0} + R_{n,1} r + R_{n,2} r^2 + R_{n,3} r^3 + R_{n,4} r^4\right),
\end{equation}
where $R_{n,i}$, $i = 0, \dots, 4$, are the coefficients given in App.\ \ref{app:coef}, the subscript $n$ is the abbreviation of $numerator$,
and the denominator takes the form,
\begin{equation}
R_d = 8 (\beta')^3 \left(d + c \alpha' \sin \theta + r^2\right)^3.
\end{equation}
Similar to the subscript $n$ in the numerator of $R$,  the subscript $d$ means $denominator$.

To solve $R = 0$, we require $R_{n,i} = 0$, resulting in five equations. 
Solving $R_{n,1} = 0$ yields two possible forms for $\beta(\theta)$:
\begin{equation}
\label{eq:sol-beta-0}
\beta(\theta) = c_1, 
\end{equation} 
\begin{equation}
\label{eq:sol-beta-0-1}
[\beta(\theta) - c_2]^2 = d + c \alpha'(\theta) \sin \theta,
\end{equation}
where $c_1$ and $c_2$ are constants of integration. The first solution, $\beta(\theta) = c_1$, leads to both the numerator and denominator of the Ricci scalar vanishing, making the ratio indeterminate; therefore, we discard it. Thus, we only keep the second solution Eq.~\eqref{eq:sol-beta-0-1}.

The equation $ R_{n,4} = 0 $ yields a nonlinear differential equation:
\begin{equation}
\label{eq:master}
-16 (\beta')^3 + 8 c \beta' \beta'' \cos \theta - 8 c \left[(\beta'')^2 + (\beta')^2 - \beta^{'''} \beta'\right] \sin \theta = 0.
\end{equation}
Additionally, the equations $ R_{n,0} = 0 $ and $ R_{n,2} = 0 $ lead to the same equation as Eq.~\eqref{eq:master} 
if $\alpha'$ is replaced by the solution in Eq.\ \eqref{eq:sol-beta-0-1}.
As a result, there are only two independent equations,\footnote{Note $R_{n,3}\equiv 0$. See App.\ \ref{app:coef}.} $R_{n,1}=0$ and $R_{n,4}=0$, among the five ones, $R_{n,i} = 0$, $i = 0, \dots, 4$. 
To solve Eq.~\eqref{eq:master}, we first note the presence of trigonometric terms 
and the absence of the function $\beta(\theta)$ itself. 
Such a property suggests a substitution to simplify the equation. 
We define the new variables,
\begin{equation}
x \coloneqq -\cos\theta, \qquad y(x) \coloneqq \beta'(x),
\end{equation}
with which we transform the original nonlinear differential equation Eq.\ \eqref{eq:master} into
\begin{equation}
y(x) \left[c \left(x^2 - 1\right) y''(x) + 2 c x y'(x) + 2 c y(x) + 2 y^2(x)\right] = c \left(x^2 - 1\right) [y'(x)]^2.
\end{equation}
To further simplify, we employ the ansatz
\begin{equation}
y(x) = \frac{\me^{-\frac{2 Y(x)}{c}}}{1 - x^2},
\end{equation}
which changes the equation into
\begin{equation}
\left(x^2 - 1\right) Y''(x) + 2 x Y'(x) + \frac{\me^{-\frac{2 Y(x)}{c}}}{x^2 - 1} = 0.
\end{equation}
Its solution is given by
\begin{equation}
\me^{-\frac{2 Y(x)}{c}} = c c_3^2 \csch^2 \left[c_3 \left(\arctanh x + c_4\right)\right],
\end{equation}
where $c_3$ and $c_4$ are constants of integration.
This solution provides a general form for $\beta'(x)$, which can now be integrated to obtain the desired transformation functions.

Now, we can derive $\beta(\theta)$ from $y = \beta'$:
\begin{equation}
\label{eq:sol-beta}
\beta(\theta) = c c_3 \coth \left[c_3 \left(\arctanh(\cos \theta) - c_4\right)\right] + c_5,
\end{equation}
where $c_5$ is a constant of integration.
Next, we determine $\alpha(\theta)$ by using the second expression in Eq.\ \eqref{eq:sol-beta-0},
\begin{equation}
\label{eq:sol-alpha}
\begin{split}
\alpha(\theta) = & \, c^{-1}\left[-d + c^2 c_3^2 + (c_2 - c_5)^2\right] \left[-\arctanh(\cos \theta) + c_4\right] \\
& + \beta(\theta) + (c_2 - c_5) \left[2 \ln (c c_3) - \ln \left(\beta^2(\theta) - 2 c_5 \beta(\theta) - c^2 c_3^2 + c_5^2\right)\right] + c_6,
\end{split}
\end{equation}
where $c_6$ is a constant of integration. 
By substituting $\alpha$ and $\beta$ back into Eq.\ \eqref{eq:metric-general}, 
we derive the general NJ class of Schwarzschild black holes. 
Notably, both $\alpha$ and $\beta$ do not depend on the parameter $b$ because the equations $R_{n,1}=0$ and $R_{n,4}=0$ do not involve $b$.

We now examine the parameters $(b, c, d)$ and constants $(c_2, c_3, c_4, c_5, c_6)$ in the solutions given by Eqs.\ \eqref{eq:sol-beta} and \eqref{eq:sol-alpha}. 
First, although $\alpha'$ and $\beta'$ rather than $\alpha$ and $\beta$ appear in the tetrad and metric, we cannot set $c_5 = c_6 = 0$. As we will demonstrate later, the non-zero values of $c_5$ and $c_6$ are essential for obtaining the correct Taub-NUT black hole.
Second, we note that Eq.\ \eqref{eq:problem} presents potential singularities at $\theta = 0$ and $\pi$ due to the presence of $\csc \theta$.
We must choose the constants in $\alpha$ and $\beta$ to eliminate any divergence in the metric at these two angles. Because $\alpha'$ depends solely on $\theta$, $\beta'$ plays a key role in addressing these singularities in $g_{22}$. Expanding $g_{22}$ around $\theta = 0$ up to the constant order, we find
\begin{equation}
g_{22} \propto 
\begin{cases}
\theta^{-2 + 2 c_3} & c_3 \ge 0, \\
\theta^{-2 (1 + c_3)} & c_3 \le 0.
\end{cases}
\end{equation}
To avoid the divergence at $\theta = 0$, the constant $c_3$ must satisfy the conditions $c_3 \ge 1$ or $c_3 \le -1$. A similar analysis shows that these conditions also apply to avoid the divergence at $\theta = \pi$.

Next, the parameters $(b, c, d)$ and constants $(c_2, c_3, c_4, c_5, c_6)$ are further constrained by the fundamental energy conditions \cite{Curiel:2014zba,Kontou:2020bta}. 
These conditions ensure that the physical interpretation of the spacetime remains consistent with the energy distribution:
\begin{itemize}
\item Weak Energy Condition (WEC): The components of Einstein’s tensor 
satisfy $-G^0_{\;0} \geq 0$ and $-G^0_{\;0} + G^i_{\;i} > 0$. This condition asserts that the total energy density, as measured by any observer following a timelike trajectory, is always non-negative.
\item Dominant Energy Condition (DEC): This condition requires $-G^0_{\;0} \geq 0$ and $-G^0_{\;0} \geq |G^i_{\;i}|$, ensuring that the energy flux remains causal and the energy-momentum propagates at or below the speed of light.
\item Strong Energy Condition (SEC): Here, $-G^0_{\;0} + \sum_i G^i_{\;i} \geq 0$ and $-G^0_{\;0} + G^i_{\;i} > 0$. This implies that the gravitational interaction tends to attract, as demonstrated by the convergence of timelike and null geodesic congruences, even when negative mass-energy densities are present.
\end{itemize}	
The energy conditions impose nonlinear inequalities on the parameters and constants. Solving these inequalities analytically is generally infeasible. Instead, they are possibly suited for verifying whether a constructed axisymmetric black hole satisfies the required energy conditions.

Let us now consider a specific example by assigning particular values to the parameters and constants. Choosing
\begin{equation}
b = c = \sqrt{d} = a, \qquad c_2 = c_5 = c_6 = 0, \qquad  c_3 = 1,  \qquad c_4 = \frac{\mi \pi}{2},
\end{equation}
where $a$ denotes the angular momentum parameter of rotating black holes, we obtain the forms of $\alpha(\theta)$ and $\beta(\theta)$ corresponding to Kerr black holes:
\begin{equation}
\alpha(\theta) = a \cos \theta, \qquad \beta(\theta) = a \cos \theta.\label{albeKerr}
\end{equation}
Alternatively, by choosing
\begin{subequations}
\begin{equation}
	b = \sqrt{a^2 - N^2}, \qquad c \to a, \qquad d \to a^2 + N^2, 
\end{equation}
\begin{equation}
c_2 = -2N, \qquad c_3 = 1, \qquad c_4 = \frac{\mi \pi}{2}, \qquad c_5 = -N, \qquad c_6 = N - \mi \pi N,
\end{equation}
\end{subequations}
we can derive the functions $\alpha(\theta)$ and $\beta(\theta)$ for Kerr-Taub-NUT black holes from Eqs.\ \eqref{eq:sol-beta} and \eqref{eq:sol-alpha}:
\begin{equation}
\alpha(\theta) = a \cos \theta + 2N \ln \sin \theta, \qquad \beta(\theta) = a \cos \theta - N,
\end{equation}
which reduce to those of pure Taub-NUT black holes when setting $a = 0$.

Our proposal differs from the treatment of Ref.~\cite{Drake:1998gf}, 
where the authors generalized the seed metric from Schwarzschild black holes 
to any spherically symmetric black hole and then adopted the NJA. Our focus is on complex transformations; 
specifically, we identify Schwarzschild black holes as the seed and correspondingly propose the general NJ transformations, 
where these transformations ensure that the resulting metric satisfies the zero-scalar-curvature condition.

\section{New  zero-scalar-curvature black hole with axisymmetry}
\label{sec:new}

In this section, we provide a new example of zero-scalar-curvature black holes derived from the Schwarzschild type using the NJA. The metric is constructed by selecting the following parameters and constants:
\begin{equation}
\label{eq:parameter_new}
b /\pp = c = \mathfrak{a}, \qquad  d = \frac{\pi^2 c^2}{4}, \qquad  c_3 = \frac{\pi}{2}, \qquad c_4 = \mi, \qquad  c_2 =c_5 = c_6 = 0,
\end{equation}
where $\mathfrak{a}$ is a parameter that is used to replace $c$ and $b\pi$.
Using these values in Eqs.\ \eqref{eq:sol-beta} and \eqref{eq:sol-alpha}, we obtain 
$\alpha(\theta)$ and $\beta(\theta)$ as follows:
\begin{equation}
	\alpha(\theta) = \beta(\theta) = \mathfrak{a} \frac{\pi}{2} \tanh\left[\frac{\pi}{2} \arctanh(\cos \theta)\right],\label{albeGerRf}
\end{equation}
which can be seen as a deformation of Eq.~\eqref{albeKerr} corresponding to Kerr black holes.
Furthermore, substituting Eq.~\eqref{albeGerRf} into
the general solutions given 
by Eq.\ \eqref{eq:metric-general}, we derive the metric components of a new zero-scalar-curvature black hole:
\begin{subequations}
\label{eq:metric_new}
\begin{equation}
g_{00} = \frac{2 \pi^2 \mathfrak{a}^2 - 4 \left[r(r - 2M) + \pi^2 \mathfrak{a}^2\right] \left[\cosh\left(\pi \arctanh(\cos \theta)\right) + 1\right]}{\left(\pi^2 \mathfrak{a}^2 + 4r^2\right) \cosh\left[\pi \arctanh(\cos \theta)\right] - \pi^2 \mathfrak{a}^2 + 4r^2},
\end{equation}
\begin{equation}
g_{11} = \frac{-\pi^2 \mathfrak{a}^2 \sech^2\left[\frac{\pi}{2} \arctanh(\cos \theta)\right] + \pi^2 \mathfrak{a}^2 + 4r^2}{4 \left[r(r - 2M) + \pi^2 \mathfrak{a}^2\right]},
\end{equation}
\begin{equation}
g_{22} = \frac{\pi^2}{16} \csc^2\theta \sech^2\left[\frac{\pi}{2} \arctanh(\cos \theta)\right] \left[\pi^2 \mathfrak{a}^2 \tanh^2\left(\frac{\pi}{2} \arctanh(\cos \theta)\right) + 4r^2\right],
\end{equation}
\begin{equation}
g_{33} = \frac{\left(\pi^3 \mathfrak{a}^2 + 4\pi r^2\right)^2 - 4\pi^4 \mathfrak{a}^2 \left[r(r - 2M) + \pi^2 \mathfrak{a}^2\right] \sech^2\left[\frac{\pi}{2} \arctanh(\cos \theta)\right]}{8 \left[\left(\pi^2 \mathfrak{a}^2 + 4r^2\right) \cosh\left(\pi \arctanh(\cos \theta)\right) - \pi^2 \mathfrak{a}^2 + 4r^2\right]},
\end{equation}
\begin{equation}
g_{03} = \frac{3 \pi^4 \mathfrak{a}^3 - 8\pi^2 M \mathfrak{a} r}{2 \left(\pi^2 \mathfrak{a}^2 + 4r^2\right) \cosh\left[\pi \arctanh(\cos \theta)\right] - 2 \pi^2 \mathfrak{a}^2 + 8r^2}.
\end{equation}
\end{subequations}
Since the general expressions Eqs.~\eqref{eq:sol-beta} and \eqref{eq:sol-alpha}  are obtained by solving the condition $R=0$, any specific parameter choice, such as in Eq.~\eqref{eq:parameter_new}, necessarily ensures that the equation of motion $R=0$ is satisfied. In particular, the resulting metric in Eq.~\eqref{eq:metric_new} must also satisfy $R=0$, which can be explicitly verified by substituting Eq.~\eqref{eq:metric_new} into the Ricci scalar.

\begin{figure}[!ht]
     \centering
     \begin{subfigure}[b]{0.45\textwidth}
         \centering
         \includegraphics[width=\textwidth]{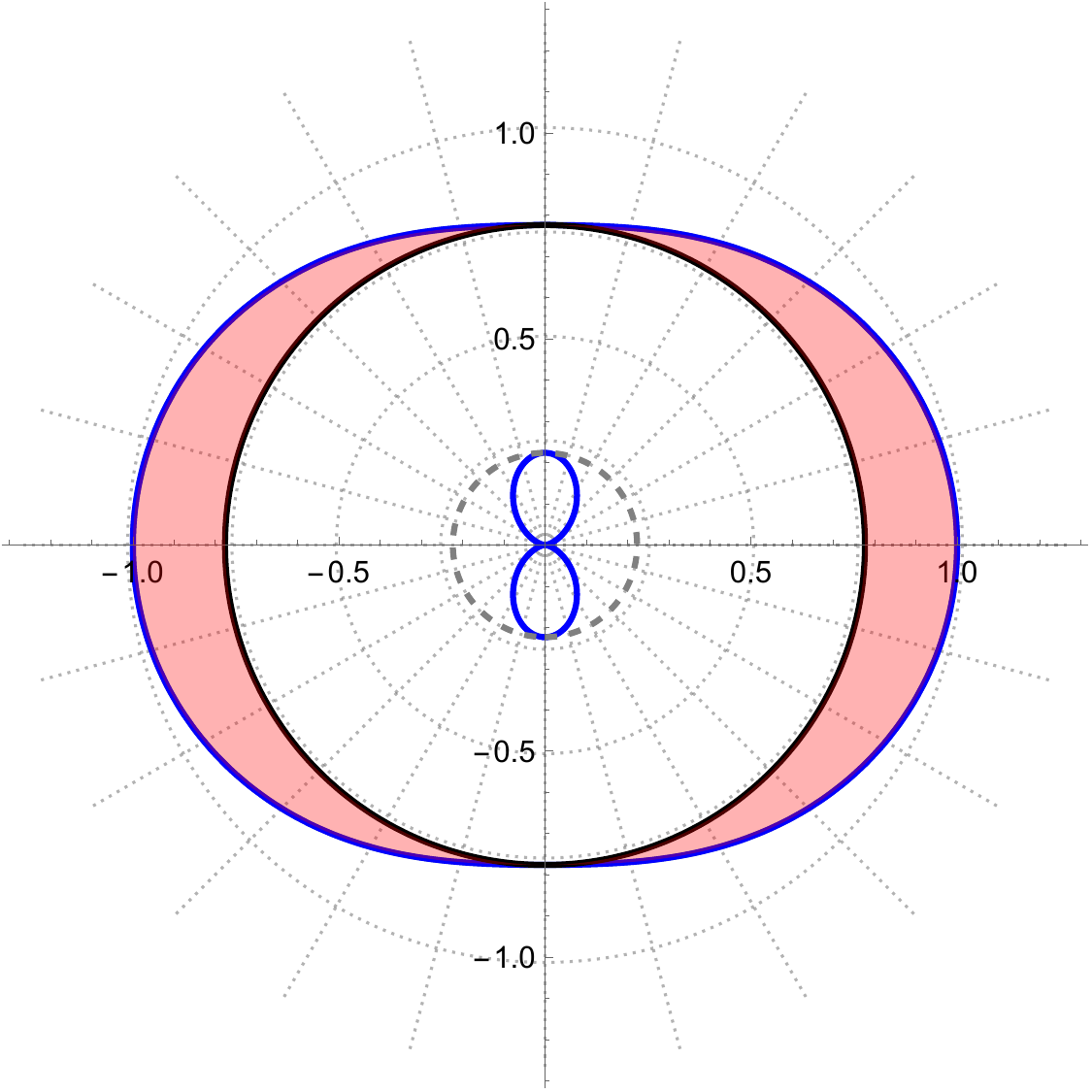}
         \caption{Kerr black holes.}
         \label{fig:kerr-structure}
     \end{subfigure}
     \begin{subfigure}[b]{0.45\textwidth}
         \centering
         \includegraphics[width=\textwidth]{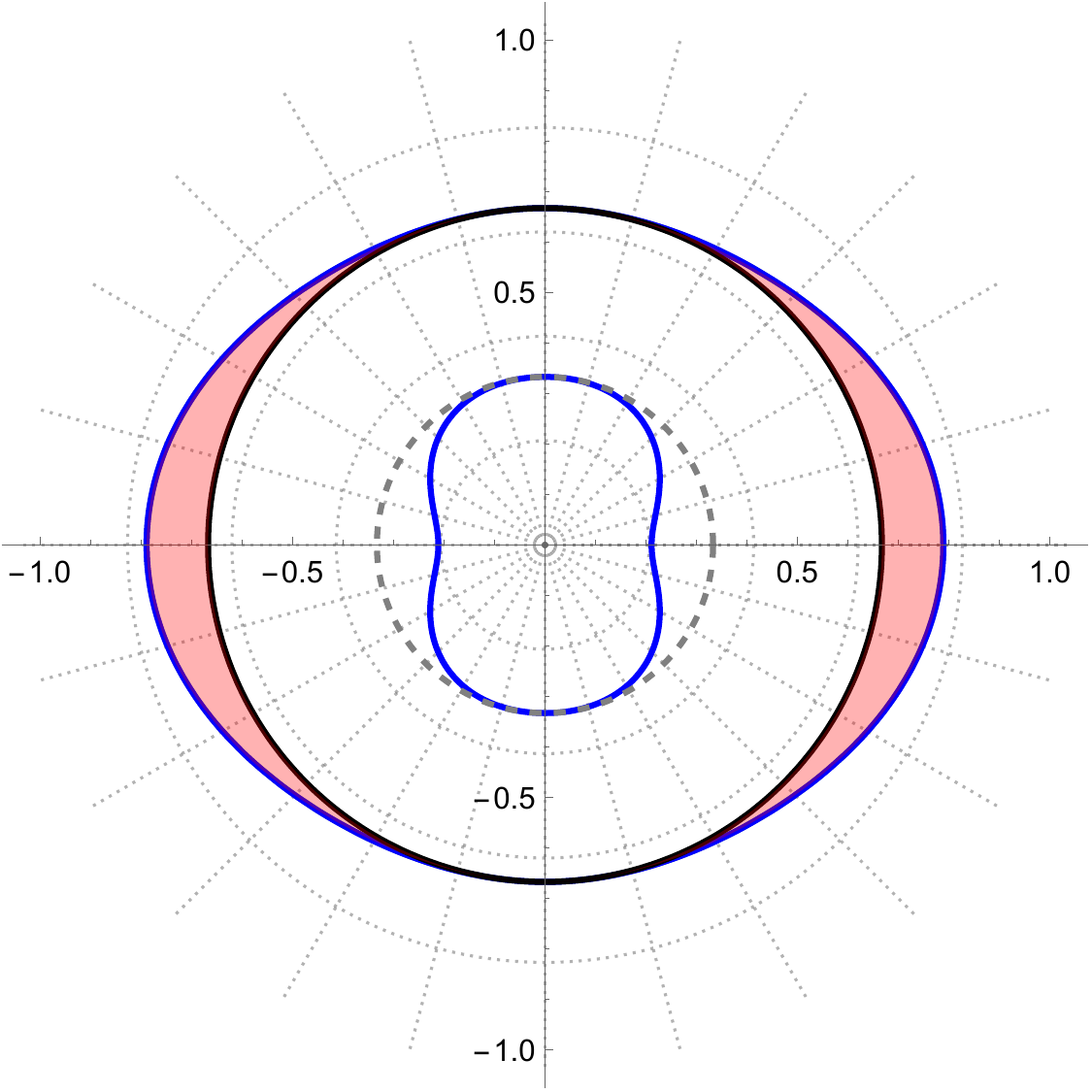}
         \caption{New black holes.}
         \label{fig:new-structure}
     \end{subfigure}
      \captionsetup{width=.9\textwidth}
       \caption{Polar slices in the Boyer–Lindquist coordinates.}
        \label{fig:compare}
\end{figure}

The horizons are determined by the  solutions of $1/g_{11} = 0$, yielding $r_\pm = M \pm \sqrt{M^2 - \pi^2 \mathfrak{a}^2}$.
The infinite redshift surfaces, characterized by the radius $r_{rs}$, are depicted by the condition $g_{00} = 0$:
\begin{equation}
2 \pi^2 \mathfrak{a}^2 - 4 \left[\pi^2 \mathfrak{a}^2 + (r_{rs} - 2M) r_{rs}\right] \left[\cosh\left(\pi \arctanh(\cos \theta)\right) + 1\right] = 0.\label{infredsurf}
\end{equation}
We compare the structure of infinite redshift surfaces, depicted by Eq.~\eqref{infredsurf}, with that of Kerr black holes in the polar slice 
using the Boyer–Lindquist coordinates, as shown in Fig.\ \ref{fig:compare}.
In this figure, the blue curves indicate the outer and inner radii of the infinite redshift surfaces, 
whereas the solid black and dashed gray curves represent the outer and inner horizons, respectively. 
The pink shadows denote the black hole ergospheres. 
Notably, the infinite redshift surfaces exhibit comparable deformations, particularly at $\theta = 0$.

To assess the existence of singularities in the spacetime, we compute the Kretschmann scalar,
\begin{equation}
\label{eq:kreshman}
K = \frac{1}{K_d}\left(K_{n,0} + K_{n,1} r + K_{n,2} r^2 + K_{n,3} r^3 + K_{n,4} r^4 + K_{n,5} r^5 + K_{n,6} r^6\right),
\end{equation}
where the coefficients $K_{n,i}$ are as provided in App.\ \eqref{app:coef-k}, and the denominator is given by
\begin{equation}
K_d = \left\{\left(\pi^2 \mathfrak{a}^2 + 4r^2\right) \cosh\left[\pi \arctanh(\cos \theta)\right] - \pi^2 \mathfrak{a}^2 + 4r^2\right\}^6.
\end{equation}
The real roots of $K_d$ occur at $r = 0$ and $\cos\theta = 0$, similar to the roots found in Kerr black holes, indicating that these black holes possess a singular loop rather than a singular point.
Moreover, the Kretschmann scalar of our solution differs from that of the known Kerr and Taub–NUT spacetimes. This enables us to conclude that the resulting black hole cannot be obtained from either Kerr or Taub–NUT through a coordinate transformation since the Kretschmann scalar is independent of the choice of coordinates.

Finally, we conclude this section with an analysis of the mechanical and thermodynamic properties of the new black hole. To compute the relevant quantities and compare them with those of the Kerr solution, we must emphasize that our analysis is conducted within Einstein’s theory of gravity. This point is important because not all thermodynamic information is determined solely by the spacetime metric; for instance, the entropy also depends on the underlying gravitational action. Next, we compare the new black hole with the Kerr black hole in  terms of the properties of Hawking temperature, Berkenstein-Hawking entropy,  and angular velocity if the two black holes have the same mass and the parameter $\mathfrak{a}$ is equivalent to the angular momentum per mass of Kerr black holes $a^{\rm Kerr}$.

Since the new axisymmetric black hole described by Eq.~\eqref{eq:metric_new} is solved within Einstein’s theory of gravity, it is expected to obey the
relation that the Hawking temperature is proportional to 
the surface gravity. That is, it is suitable to compare the surface gravity of the two black holes. 
The surface gravity of the new black hole can be computed from the following relation:
\begin{equation}
	\kappa^2=-\frac{1}{2}\nabla^\mu \xi^\nu \nabla_\mu \xi_\nu \Big|_{\rm horizon}, \qquad \xi^\nu = (1,0,0,\Omega),
\end{equation}
where $\Omega$ is the angular velocity given by Eq.~\eqref{newangvel}. Thus, we obtain
\begin{equation}
	\kappa = \frac{r_+-M}{r_+^2+\frac{\pp^2 \mathfrak{a}^2}{4}}.
\end{equation}
For the Kerr black hole, its surface gravity is~\cite{Zhao:2016irg,Chen:2022igr} 
\begin{equation}
\kappa_{\rm Kerr}=\frac{r_+^{\rm Kerr}-M}{(r_+^{\rm Kerr})^2+(a^{\rm Kerr})^2}.
\end{equation}
It is obvious that $\kappa\neq \kappa_{\rm Kerr}$ even if the two black holes have the same mass and $\mathfrak{a}=a^{\rm Kerr}$ \footnote{
Scaling $\mathfrak{a}$ by the replacement $\pp\mathfrak{a}/2 \to \mathfrak{a}$ also does not lead to $\kappa = \kappa^{\rm Kerr}$. Under this transformation, the outer horizon becomes
$r_+ = M + \sqrt{M^2 - 4\mathfrak{a}^2}$,
which clearly differs from the Kerr horizon $r_+^{\rm Kerr}$. Furthermore, such a scaling does not transform metric \eqref{eq:metric_new} into the Kerr metric, and therefore cannot establish equivalence with the Kerr spacetime.
}.

Next, we consider the Bekenstein–Hawking entropy. According to the Bekenstein–Hawking area law, the entropy is proportional to~\cite{Lan:2023eqt} the horizon area, $S = A/4$. We can compute the horizon area from Eq.~\eqref{eq:metric_new}:
\begin{equation}
	A =  2 \pp^2 \left(r_+^2 +\frac{\pp^2 \mathfrak{a}^2}{4}\right)
    \label{horareanew}
\end{equation}
by integrating the surface with $t=\text{const.}$ and $r=r_+$. For comparison, the horizon area of Kerr black holes takes the form 
\begin{equation}
A_{\rm Kerr}=4\pp[(r^{\rm Kerr}_+)^2+(a^{\rm Kerr})^2], 
\end{equation}
which is evidently different from Eq.~\eqref{horareanew}.

The angular velocity at the event horizon can be obtained~\cite{Zhao:2016irg,Chen:2022igr} from the tangent null vector $l^\alpha=(1,0,0,\Omega)$, yielding
\begin{equation}
		\Omega=\frac{\mathfrak{a}}{r_+^2+\frac{\pp^2 \mathfrak{a}^2}{4}},\label{newangvel}
\end{equation}
which represents the angular velocity of light rays relative to an observer at infinity and can be interpreted as the black hole’s angular velocity. For comparison, the angular velocity of a Kerr black hole has the form
\begin{equation}	
\Omega_{\rm Kerr}=\frac{a^{\rm Kerr}}{2M r^{\rm Kerr}_+},
\end{equation}
which is also different from Eq.~\eqref{newangvel}.

\section{Conclusion}
\label{sec:conclusion}

In this study, we aim to construct axisymmetric black holes that satisfy the zero-scalar-curvature condition 
and can be generated by the NJA from Schwarzschild black holes as the seed metric. 
We successfully derive the general forms of complex transformations and the corresponding axisymmetric metrics. 
Our results include  Kerr, Taub-NUT, and Kerr-Taub-NUT black holes as special cases
for different choices of parameters. Additionally, our findings indicate the existence of 
new axisymmetric black holes, which we designate as the NJ class of Schwarzschild black holes.

The methodology developed in this study may be extended to related areas. 
For instance, it can be applied to construct axisymmetric black holes in the Chern-Simons gravity. 
This would involve inverting the specific form of complex transformations 
when the vanishing Pontryagin density is imposed as a constraint. 
Another possible application is the construction of regular black holes \cite{Lan:2023cvz} that satisfy the relevant equations of motion. Notably, a class of regular black holes exists
that does not originate from quantum corrections, such as the ABG black holes \cite{Ayon-Beato:1998hmi}, 
the solutions to gravity coupled to nonlinear electrodynamics, 
where the corresponding equations of motion should be imposed for searching  axisymmetric black holes besides the zero-scalar-curvature condition.
Moreover, for the regular black holes derived from quantum corrections \cite{Lewandowski:2022zce,Zhang:2024khj,Shi:2024vki,Lin:2024flv}, 
our method may facilitate the generation of axisymmetric solutions that meet specific conditions.
This will be the focus of our future research.

\section*{Acknowledgement}

This work was supported in part by the National Natural Science Foundation of China under Grant No.\ 12175108. L.C. is also supported by Yantai University under Grant No.\ WL22B224.

\appendix

\section{Coefficients in Eq.~\eqref{eq:ricci}}
\label{app:coef}

Here, we present the coefficients of the numerator in Eq.\  \eqref{eq:ricci}.

\begin{equation}
\begin{split}
R_{n,0}  =& -4 \Big\{
	2 a^2 b^2 c  (\beta '' )^2\sin(\theta)  -b c (\beta ' ) 
	\Big[2 a^2 b \left(\beta ^{(3)} \sin(\theta) + \beta ''\cos(\theta) \right)\\
	&-2 a c \sin ^2\theta  \alpha '' \beta ''+b c \sin ^2\theta  (\alpha '' )^2\Big]
	+4 a b^2 (a+b) (\beta ' )^3\\
	&+c (\beta ' )^2 \left[2 a b \sin \theta  \left(a b-c \alpha ^{(3)} \sin \theta \right)
	-2 a b c \sin (2 \theta ) \alpha ''+c^2 \sin ^3\theta  (\alpha '' )^2\right]
	\Big\}\\
	&+8 c \sin \theta  \alpha '
	\Big\{c \beta '\Big[a b \left(2 \beta ^{(3)} \sin \theta +\cos \theta  \beta ''\right)
	+\alpha '' \left(b^2 \cos \theta -c \sin ^2\theta  \beta ''\right)\Big]\\
	&-2 b (2 a+b) (\beta ')^3+c \sin \theta  (\beta ')^2 \left[-3 a b+c \alpha ^{(3)} \sin \theta 
	+c \cos \theta  \alpha ''\right]\\
	&-2 a b c \sin \theta  (\beta '')^2\Big\}
	+c^2 (\alpha ' )^2 \Big\{4 \beta ' \left[b^2 \cos ^2\theta +2 c \beta ^{(3)} \sin ^3\theta \right]-16 \sin ^2\theta  (\beta ' )^3\\
	&-8 c \sin ^3\theta  (\beta '')^2
	+c [3 \sin (3 \theta )-13 \sin \theta ] (\beta ' )^2\Big\},
\end{split}
\end{equation}

\begin{equation}
\begin{split}
R_{n,1}=-8 M \beta ' \Big\{
&c \Big[
2 \sin \theta  \alpha ' \left(c \cos \theta  \alpha ''-2 (\beta ' )^2\right)\\
&+c \sin ^2\theta  (\alpha '')^2
+c \cos ^2\theta  (\alpha ')^2
\Big]
-4 a b (\beta ')^2
\Big\},
\end{split}
\end{equation}

\begin{equation}
\begin{split}
R_{n,2} = 
&4 \Big\{
-12 a b (\beta ')^3
+c \beta' \left[4 a b  \beta '' \cos(\theta) 
+\sin \theta  \left(4 a b \beta ^{(3)}
+c \sin \theta  \alpha '' \left(\alpha ''
-2 \beta ''\right)\right)\right]\\
&+2 c \sin \theta  (\beta ')^2 
\left[-2 a b+c \alpha ^{(3)} \sin \theta +2 c \cos \theta  \alpha '' \right]\\
&-4 a b c \sin \theta  (\beta '' )^2+c^2 \cos ^2\theta  (\alpha ' )^2 \beta '
+2 c \sin \theta  \alpha '
\Big[
	-6 (\beta ' )^3\\
	&+c \cos \theta  \beta ' \left(\alpha ''
	+\beta ''\right)
	+c \sin \theta  \left(-2 (\beta '' )^2-3 (\beta ')^2
	+2 \beta ^{(3)} \beta '\right)
\Big]-4 c \sin \theta  (\beta ')^4
\Big\},
\end{split}
\end{equation}

\begin{equation}
R_{n,3}=0,
\end{equation}

\begin{equation}
R_{n,4} = -16 (\beta ')^3
+8 c \beta ' \beta ''\cos(\theta)
-8 c \left[(\beta '')^2+(\beta ' )^2-\beta ^{(3)} \beta '\right]
\sin (\theta).
\end{equation}

\section{Coefficients in Eq.~\eqref{eq:kreshman}}
\label{app:coef-k}

Here, we display the coefficients of the numerator in Eq.\  \eqref{eq:kreshman}.

\begin{equation}
\begin{split}
K_{n,0}= 12288 \pp ^6 \mathfrak{a}^6 
& \sinh ^4\left[\frac{ \pp }{2} \arctanh(\cos \theta)\right] 
\cosh ^6\left[\frac{ \pp }{2} \arctanh(\cos \theta)\right] \\
& \times\left[\left(21 \pp ^2 \mathfrak{a}^2-8 M^2\right) \cosh \left(\pp  \arctanh(\cos \theta)\right)+8 M^2+21 \pp ^2 \mathfrak{a}^2\right],
\end{split}
\end{equation}
\begin{equation}
K_{n,1}=-5898240 \pp ^6 M \mathfrak{a}^6 \sinh ^4\left[\frac{ \pp }{2} \arctanh(\cos \theta)\right] 
\cosh ^8\left[\frac{ \pp }{2} \arctanh(\cos \theta)\right],
\end{equation}
\begin{equation}
\begin{split}
K_{n,2}=  -294912 \pp ^4 \mathfrak{a}^4 & \sinh ^2\left[\frac{ \pp }{2} \arctanh(\cos \theta)\right] 
\cosh ^8\left[\frac{ \pp }{2} \arctanh(\cos \theta)\right]  \\
& \times\left[\left(17 \pp ^2 \mathfrak{a}^2-20 M^2\right) \cosh \left(\pp  \arctanh(\cos \theta)\right)+20 M^2+17 \pp ^2 \mathfrak{a}^2\right],
\end{split}
\end{equation}
\begin{equation}
K_{n,3}=47185920 \pp ^4 M \mathfrak{a}^4 \sinh ^2\left[\frac{ \pp }{2} \arctanh(\cos \theta)\right] \cosh ^{10}\left[\frac{ \pp }{2} \arctanh(\cos \theta)\right],
\end{equation}
\begin{equation}
\begin{split}
K_{n,4}  =589824 \pp ^2 & \mathfrak{a}^2   \cosh ^{10}\left[\frac{ \pp }{2} \arctanh(\cos \theta)\right]\\
& \times \left[\left(7 \pp ^2 \mathfrak{a}^2-40 M^2\right) \cosh \left(\pp  \arctanh(\cos \theta)\right)+40 M^2+7 \pp ^2 \mathfrak{a}^2\right],
 \end{split}
\end{equation}
\begin{equation}
K_{n,5}=-18874368 \pp ^2 M \mathfrak{a}^2 \cosh ^{12}\left[\frac{ \pp }{2} \arctanh(\cos \theta)\right],
\end{equation}
\begin{equation}
K_{n,6}=12582912 M^2 \cosh ^{12}\left[\frac{ \pp }{2} \arctanh(\cos \theta)\right].
\end{equation}

\bibliographystyle{utphys}
\bibliography{references}

\providecommand{\href}[2]{#2}\begingroup\raggedright\begin{thebibliography}{10}

\bibitem{Newman:1965tw}
E.~T. Newman and A.~I. Janis, ``{Note on the Kerr spinning particle metric},''
  \href{http://dx.doi.org/10.1063/1.1704350}{{\em J. Math. Phys.} {\bfseries 6}
  (1965) 915--917}.

\bibitem{Erbin:2016lzq}
H.~Erbin, ``{Janis-Newman algorithm: generating rotating and NUT charged black
  holes},'' \href{http://dx.doi.org/10.3390/universe3010019}{{\em Universe}
  {\bfseries 3} no.~1, (2017) 19},
  \href{http://arxiv.org/abs/1701.00037}{{\ttfamily arXiv:1701.00037 [gr-qc]}}.

\bibitem{Alexander:2009tp}
S.~Alexander and N.~Yunes, ``{Chern-Simons Modified General Relativity},''
  \href{http://dx.doi.org/10.1016/j.physrep.2009.07.002}{{\em Phys. Rept.}
  {\bfseries 480} (2009) 1--55},
  \href{http://arxiv.org/abs/0907.2562}{{\ttfamily arXiv:0907.2562 [hep-th]}}.

\bibitem{Newman:1961qr}
E.~Newman and R.~Penrose, ``{An Approach to gravitational radiation by a method
  of spin coefficients},'' \href{http://dx.doi.org/10.1063/1.1724257}{{\em J.
  Math. Phys.} {\bfseries 3} (1962) 566--578}.

\bibitem{Erbin:2014aya}
H.~Erbin, ``{Janis\textendash{}Newman algorithm: simplifications and gauge
  field transformation},''
  \href{http://dx.doi.org/10.1007/s10714-015-1860-1}{{\em Gen. Rel. Grav.}
  {\bfseries 47} (2015) 19}, \href{http://arxiv.org/abs/1410.2602}{{\ttfamily
  arXiv:1410.2602 [gr-qc]}}.

\bibitem{Azreg-Ainou:2014pra}
M.~Azreg-A\"\i{}nou, ``{Generating rotating regular black hole solutions
  without complexification},''
  \href{http://dx.doi.org/10.1103/PhysRevD.90.064041}{{\em Phys. Rev. D}
  {\bfseries 90} no.~6, (2014) 064041},
  \href{http://arxiv.org/abs/1405.2569}{{\ttfamily arXiv:1405.2569 [gr-qc]}}.

\bibitem{Newman:1965my}
E.~T. Newman, R.~Couch, K.~Chinnapared, A.~Exton, A.~Prakash, and R.~Torrence,
  ``{Metric of a Rotating, Charged Mass},''
  \href{http://dx.doi.org/10.1063/1.1704351}{{\em J. Math. Phys.} {\bfseries 6}
  (1965) 918--919}.

\bibitem{Adamo:2014baa}
T.~Adamo and E.~T. Newman, ``{The Kerr-Newman metric: A Review},''
  \href{http://dx.doi.org/10.4249/scholarpedia.31791}{{\em Scholarpedia}
  {\bfseries 9} (2014) 31791}, \href{http://arxiv.org/abs/1410.6626}{{\ttfamily
  arXiv:1410.6626 [gr-qc]}}.

\bibitem{Johannsen:2011dh}
T.~Johannsen and D.~Psaltis, ``{A Metric for Rapidly Spinning Black Holes
  Suitable for Strong-Field Tests of the No-Hair Theorem},''
  \href{http://dx.doi.org/10.1103/PhysRevD.83.124015}{{\em Phys. Rev. D}
  {\bfseries 83} (2011) 124015},
  \href{http://arxiv.org/abs/1105.3191}{{\ttfamily arXiv:1105.3191 [gr-qc]}}.

\bibitem{Bambi:2015kza}
C.~Bambi, ``{Testing black hole candidates with electromagnetic radiation},''
  \href{http://dx.doi.org/10.1103/RevModPhys.89.025001}{{\em Rev. Mod. Phys.}
  {\bfseries 89} no.~2, (2017) 025001},
  \href{http://arxiv.org/abs/1509.03884}{{\ttfamily arXiv:1509.03884 [gr-qc]}}.

\bibitem{Konoplya:2016jvv}
R.~Konoplya, L.~Rezzolla, and A.~Zhidenko, ``{General parametrization of
  axisymmetric black holes in metric theories of gravity},''
  \href{http://dx.doi.org/10.1103/PhysRevD.93.064015}{{\em Phys. Rev. D}
  {\bfseries 93} no.~6, (2016) 064015},
  \href{http://arxiv.org/abs/1602.02378}{{\ttfamily arXiv:1602.02378 [gr-qc]}}.

\bibitem{Arkani-Hamed:2019ymq}
N.~Arkani-Hamed, Y.-t. Huang, and D.~O'Connell, ``{Kerr black holes as
  elementary particles},''
  \href{http://dx.doi.org/10.1007/JHEP01(2020)046}{{\em JHEP} {\bfseries 01}
  (2020) 046}, \href{http://arxiv.org/abs/1906.10100}{{\ttfamily
  arXiv:1906.10100 [hep-th]}}.

\bibitem{Perlick:2021aok}
V.~Perlick and O.~Y. Tsupko, ``{Calculating black hole shadows: Review of
  analytical studies},''
  \href{http://dx.doi.org/10.1016/j.physrep.2021.10.004}{{\em Phys. Rept.}
  {\bfseries 947} (2022) 1--39},
  \href{http://arxiv.org/abs/2105.07101}{{\ttfamily arXiv:2105.07101 [gr-qc]}}.

\bibitem{Modesto:2010rv}
L.~Modesto and P.~Nicolini, ``{Charged rotating noncommutative black holes},''
  \href{http://dx.doi.org/10.1103/PhysRevD.82.104035}{{\em Phys. Rev. D}
  {\bfseries 82} (2010) 104035},
  \href{http://arxiv.org/abs/1005.5605}{{\ttfamily arXiv:1005.5605 [gr-qc]}}.

\bibitem{Bambi:2013ufa}
C.~Bambi and L.~Modesto, ``{Rotating regular black holes},''
  \href{http://dx.doi.org/10.1016/j.physletb.2013.03.025}{{\em Phys. Lett. B}
  {\bfseries 721} (2013) 329--334},
  \href{http://arxiv.org/abs/1302.6075}{{\ttfamily arXiv:1302.6075 [gr-qc]}}.

\bibitem{Toshmatov:2017zpr}
B.~Toshmatov, Z.~Stuchl\'\i{}k, and B.~Ahmedov, ``{Generic rotating regular
  black holes in general relativity coupled to nonlinear electrodynamics},''
  \href{http://dx.doi.org/10.1103/PhysRevD.95.084037}{{\em Phys. Rev. D}
  {\bfseries 95} no.~8, (2017) 084037},
  \href{http://arxiv.org/abs/1704.07300}{{\ttfamily arXiv:1704.07300 [gr-qc]}}.

\bibitem{Kamenshchik:2023woo}
A.~Kamenshchik and P.~Petriakova, ``{Newman-Janis algorithm\textquoteright{}s
  application to regular black hole models},''
  \href{http://dx.doi.org/10.1103/PhysRevD.107.124020}{{\em Phys. Rev. D}
  {\bfseries 107} no.~12, (2023) 124020},
  \href{http://arxiv.org/abs/2305.04697}{{\ttfamily arXiv:2305.04697 [gr-qc]}}.

\bibitem{Drake:1998gf}
S.~P. Drake and P.~Szekeres, ``{Uniqueness of the Newman-Janis algorithm in
  generating the Kerr-Newman metric},''
  \href{http://dx.doi.org/10.1023/A:1001920232180}{{\em Gen. Rel. Grav.}
  {\bfseries 32} (2000) 445--458},
  \href{http://arxiv.org/abs/gr-qc/9807001}{{\ttfamily arXiv:gr-qc/9807001}}.

\bibitem{Beltracchi:2021ris}
P.~Beltracchi and P.~Gondolo, ``{Physical interpretation of Newman-Janis
  rotating systems. I. A unique family of Kerr-Schild systems},''
  \href{http://dx.doi.org/10.1103/PhysRevD.104.124066}{{\em Phys. Rev. D}
  {\bfseries 104} no.~12, (2021) 124066},
  \href{http://arxiv.org/abs/2104.02255}{{\ttfamily arXiv:2104.02255 [gr-qc]}}.

\bibitem{Lan:2023vaa}
C.~Lan, M.-H. Li, and Y.-G. Miao, ``{Phase diagrams of quasinormal frequencies
  for Schwarzschild, Kerr, and Taub-NUT black holes},''
  \href{http://dx.doi.org/10.1103/PhysRevD.110.044045}{{\em Phys. Rev. D}
  {\bfseries 110} no.~4, (2024) 044045},
  \href{http://arxiv.org/abs/2312.05457}{{\ttfamily arXiv:2312.05457 [gr-qc]}}.

\bibitem{Gurses:1975vu}
M.~Gurses and G.~Feza, ``{Lorentz Covariant Treatment of the Kerr-Schild
  Metric},'' \href{http://dx.doi.org/10.1063/1.522480}{{\em J. Math. Phys.}
  {\bfseries 16} (1975) 2385}.

\bibitem{Dymnikova:2015hka}
I.~Dymnikova and E.~Galaktionov, ``{Regular rotating electrically charged black
  holes and solitons in non-linear electrodynamics minimally coupled to
  gravity},'' \href{http://dx.doi.org/10.1088/0264-9381/32/16/165015}{{\em
  Class. Quant. Grav.} {\bfseries 32} no.~16, (2015) 165015},
  \href{http://arxiv.org/abs/1510.01353}{{\ttfamily arXiv:1510.01353 [gr-qc]}}.

\bibitem{Erbin:2014lwa}
H.~Erbin and L.~Heurtier, ``{Five-dimensional Janis\textendash{}Newman
  algorithm},'' \href{http://dx.doi.org/10.1088/0264-9381/32/16/165004}{{\em
  Class. Quant. Grav.} {\bfseries 32} no.~16, (2015) 165004},
  \href{http://arxiv.org/abs/1411.2030}{{\ttfamily arXiv:1411.2030 [gr-qc]}}.

\bibitem{Griffiths:2009dfa}
J.~B. Griffiths and J.~Podolsky,
  \href{http://dx.doi.org/10.1017/CBO9780511635397}{{\em {Exact Space-Times in
  Einstein's General Relativity}}}.
\newblock Cambridge Monographs on Mathematical Physics. Cambridge University
  Press, Cambridge, 2009.

\bibitem{Curiel:2014zba}
E.~Curiel, ``{A Primer on Energy Conditions},''
  \href{http://dx.doi.org/10.1007/978-1-4939-3210-8_3}{{\em Einstein Stud.}
  {\bfseries 13} (2017) 43--104},
  \href{http://arxiv.org/abs/1405.0403}{{\ttfamily arXiv:1405.0403
  [physics.hist-ph]}}.

\bibitem{Kontou:2020bta}
E.-A. Kontou and K.~Sanders, ``{Energy conditions in general relativity and
  quantum field theory},''
  \href{http://dx.doi.org/10.1088/1361-6382/ab8fcf}{{\em Class. Quant. Grav.}
  {\bfseries 37} no.~19, (2020) 193001},
  \href{http://arxiv.org/abs/2003.01815}{{\ttfamily arXiv:2003.01815 [gr-qc]}}.

\bibitem{Zhao:2016irg}
L.~Zhao, {\em {Introduction to Relativity and Gravity (in Chinese)}}.
\newblock 21st Century Series on Frontiers of Theoretical Physics and Its
  Interdisciplinary Fields. Science Press, Beijing, 2016.11.

\bibitem{Chen:2022igr}
B.~Chen, {\em {Introduction to General Relativity (in Chinese)}}.
\newblock 21st Century Physics Planning Textbooks · Basic Course Series.
  Beijing University Press, Beijing, 2022.4.

\bibitem{Lan:2023eqt}
C.~Lan and Y.-G. Miao, ``{Entropy of Regular Black Holes in
  Einstein{\textquoteright}s Gravity},''
  \href{http://dx.doi.org/10.1088/0256-307X/40/12/120401}{{\em Chin. Phys.
  Lett.} {\bfseries 40} no.~12, (2023) 120401}.

\bibitem{Lan:2023cvz}
C.~Lan, H.~Yang, Y.~Guo, and Y.-G. Miao, ``{Regular Black Holes: A Short Topic
  Review},'' \href{http://dx.doi.org/10.1007/s10773-023-05454-1}{{\em Int. J.
  Theor. Phys.} {\bfseries 62} no.~9, (2023) 202},
  \href{http://arxiv.org/abs/2303.11696}{{\ttfamily arXiv:2303.11696 [gr-qc]}}.

\bibitem{Ayon-Beato:1998hmi}
E.~Ayon-Beato and A.~Garcia, ``{Regular black hole in general relativity
  coupled to nonlinear electrodynamics},''
  \href{http://dx.doi.org/10.1103/PhysRevLett.80.5056}{{\em Phys. Rev. Lett.}
  {\bfseries 80} (1998) 5056--5059},
  \href{http://arxiv.org/abs/gr-qc/9911046}{{\ttfamily arXiv:gr-qc/9911046}}.

\bibitem{Lewandowski:2022zce}
J.~Lewandowski, Y.~Ma, J.~Yang, and C.~Zhang, ``{Quantum Oppenheimer-Snyder and
  Swiss Cheese Models},''
  \href{http://dx.doi.org/10.1103/PhysRevLett.130.101501}{{\em Phys. Rev.
  Lett.} {\bfseries 130} no.~10, (2023) 101501},
  \href{http://arxiv.org/abs/2210.02253}{{\ttfamily arXiv:2210.02253 [gr-qc]}}.

\bibitem{Zhang:2024khj}
C.~Zhang, J.~Lewandowski, Y.~Ma, and J.~Yang, ``{Black Holes and Covariance in
  Effective Quantum Gravity},''
  \href{http://arxiv.org/abs/2407.10168}{{\ttfamily arXiv:2407.10168 [gr-qc]}}.

\bibitem{Shi:2024vki}
Z.~Shi, X.~Zhang, and Y.~Ma, ``{Higher-dimensional quantum Oppenheimer-Snyder
  model},'' \href{http://dx.doi.org/10.1103/PhysRevD.110.104074}{{\em Phys.
  Rev. D} {\bfseries 110} no.~10, (2024) 104074},
  \href{http://arxiv.org/abs/2408.15821}{{\ttfamily arXiv:2408.15821 [gr-qc]}}.

\bibitem{Lin:2024flv}
J.~Lin and X.~Zhang, ``{Effective four-dimensional loop quantum black hole with
  a cosmological constant},''
  \href{http://dx.doi.org/10.1103/PhysRevD.110.026002}{{\em Phys. Rev. D}
  {\bfseries 110} no.~2, (2024) 026002},
  \href{http://arxiv.org/abs/2402.13638}{{\ttfamily arXiv:2402.13638 [gr-qc]}}.

\end{thebibliography}\endgroup
\end{document}